# A Decentralized Academic Certificate Issuance System Using Smart Contracts on the Tron Network

Ana Julia Evangelista Andrade and Flavio Cezar Amate

Department of Informatics, Federal Institute of Education, Science and Technology of São Paulo (IFSP), Bragança Paulista, SP, Brazil

## ABSTRACT

*This paper presents the design, implementation, and evaluation of a decentralized system for issuing and verifying academic certificates based on blockchain technology. The proposed solution addresses common limitations of traditional certification models, such as susceptibility to forgery, reliance on centralized infrastructures, and inefficient verification processes. The system is built on the TRON blockchain and integrates smart contracts written in Solidity, a decentralized web application (dApp) for user interaction, and the InterPlanetary File System (IPFS) for decentralized storage of certificate metadata. The methodology comprised architectural design, smart contract development, and the implementation of a web-based interface, followed by functional, security, performance, and usability evaluations. Experimental results show that the system correctly supports certificate issuance and public verification, enforces access control, and resists common misuse scenarios. Performance analysis indicates low confirmation latency and negligible transaction costs, making the solution suitable for large-scale academic environments. Additionally, usability assessment using the System Usability Scale (SUS) resulted in a score of 76.67, indicating good user acceptance. Overall, the results demonstrate the technical feasibility and practical viability of the proposed approach, highlighting the TRON blockchain as an effective and cost-efficient infrastructure for decentralized academic certification systems.*

## 1. INTRODUCTION

Academic certification plays a fundamental role in validating qualifications, skills, and educational achievements, being widely used by educational institutions, students, employers, and regulatory bodies. However, traditional models for issuing and verifying academic certificates, typically based on physical documents or centralized digital systems, present significant limitations, including vulnerability to forgery, bureaucratic processes, low interoperability between institutions, and strong dependence on centralized infrastructures and intermediaries for authenticity validation [1], [2]. These limitations make verification processes slow, costly, and, in many cases, unreliable.

In recent years, blockchain technology has emerged as a promising alternative to address these challenges. By design, blockchain systems provide properties such as immutability, decentralization, transparency, and auditability, enabling secure and verifiable data registration without reliance on a central authority [3]. These characteristics have motivated increasing interest in applying blockchain-based solutions to the educational domain, particularly for the issuance and verification of academic certificates and diplomas.

Several studies have explored the use of blockchain for academic certification. Ataşen and Aslan proposed a decentralized digital certification platform in which academic credentials are represented by cryptographic hashes registered on the blockchain, reducing reliance on centralized databases and mitigating forgery risks [4]. Similarly, Gaikwad et al. developed a blockchain-based system using smart contracts to enable fast and reliable verification of academic certificates through web-based applications [5]. Ojog et al. investigated the adoption of decentralized applications (DApps) and smart contracts in educational environments, highlighting their potential to support decentralized management and verification of academic records, credentials, and certifications [6].

Other works emphasize the role of blockchain as a mechanism for enhancing security and trust in academic credential management. Fartitchou et al. introduced BlockMEDC, a smart contract–based framework that automates diploma issuance and reduces administrative costs while improving transparency [8]. Additional research indicates that blockchain-based systems for academic certificate verification can significantly benefit recruitment processes, allowing employers and other stakeholders to independently verify credentials with greater reliability and reduced dependency on central authorities [9], [10].

Despite these advances, many blockchain-based certification solutions rely on public networks such as Ethereum, which, although widely adopted, may present limitations related to transaction costs, scalability, and confirmation latency when applied to large-scale systems. Several studies report that public blockchains often suffer from low throughput and high latency under increasing transaction loads, which can constrain their suitability for high-frequency and large-scale applications [11], [12].

In this context, the TRON blockchain network stands out due to its high throughput, reduced transaction costs, and low confirmation latency, as well as its compatibility with smart contracts written in Solidity and seamless integration with decentralized web applications [13], [14]. Such characteristics make TRON a suitable candidate for blockchain-based academic certification systems that demand efficiency, scalability, and cost-effectiveness.

Motivated by this gap, this paper presents the design, implementation, and evaluation of a decentralized system for issuing and verifying academic certificates based on the TRON blockchain. The proposed solution employs smart contracts to ensure immutable certificate registration, a decentralized web application (dApp) for user interaction, and the InterPlanetary File System (IPFS) for decentralized storage of certificate metadata [15]. Authentication and transaction signing are handled through the TronLink wallet, ensuring that only authorized entities can issue certificates.

The main contributions of this work are: (i) the proposal of a complete decentralized architecture for academic certification using the TRON blockchain; (ii) the implementation of smart contracts for certificate registration and validation; (iii) the integration of blockchain, decentralized storage, and web-based interfaces; and (iv) an experimental evaluation of the system through functional, security, performance, and usability tests.

## 2. METHODOLOGY

This work is characterized as applied research with an experimental approach, aiming at the development and evaluation of a decentralized system for issuing and verifying academic certificates using blockchain technology. The adopted methodology involved the definition of the system architecture, smart contract development, the implementation of a decentralized web application, and the execution of functional, security, performance, and usability tests, considering established practices and solutions reported in the literature.

## 2.1. Tools and Technologies

Based on the theoretical study, the main tools and technologies used in the system development were selected. The TRON blockchain was adopted as the primary infrastructure due to its high scalability, low transaction costs, and compatibility with smart contracts written in Solidity. Interaction with the blockchain was performed using the TronWeb JavaScript library, while transaction authentication and signing were handled through the TronLink digital wallet.

Smart contract development was carried out using the Solidity programming language and specific tools from the TRON ecosystem. The user interface was implemented as a decentralized web application (dApp), developed using the React.js library. For decentralized storage of certificate metadata, the InterPlanetary File System (IPFS) was employed, ensuring data availability and persistence without reliance on centralized servers.

## 2.2. Requirements Definition

The definition of system requirements was carried out based on problem analysis and the literature review. The functional requirements encompassed the decentralized issuance, registration, retrieval, and validation of academic certificates. Non-functional requirements included aspects related to security, data integrity, usability, scalability, record immutability, and overall system performance.

## 2.3. Smart Contract Development

Smart contracts responsible for certificate registration and validation were implemented in Solidity and deployed initially on the TRON test network. Each certificate is represented by a cryptographic hash generated from its attributes, ensuring uniqueness and preventing duplication. Access control mechanisms were implemented to ensure that only authorized addresses could issue certificates.

## 2.4. Decentralized Web Application

The decentralized web application was developed to enable intuitive interaction between users and smart contracts. The interface allows authorized entities to issue academic certificates and enables public validation of records by any interested party. Integration between the frontend and the TRON blockchain was achieved using the TronWeb library, with authentication provided by the TronLink wallet.

The system workflow involves the submission of certificate data through the web application, the generation of the corresponding cryptographic hash, the registration of this information on the blockchain, and the storage of metadata on IPFS. This process ensures that certificates can be verified in a decentralized, transparent, and secure manner, as illustrated in Figure 1.

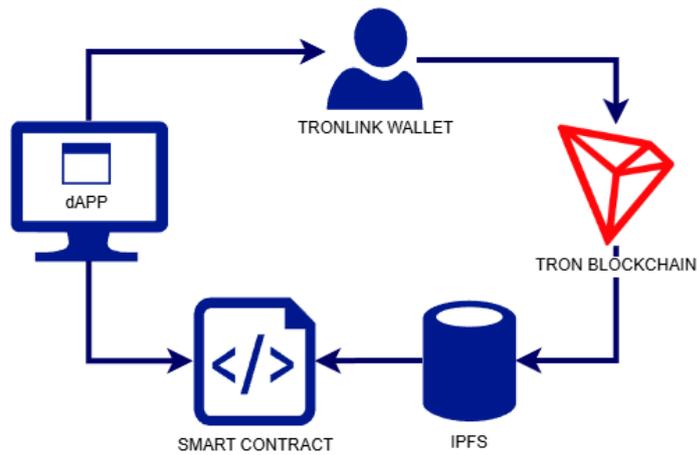

Figure 1. Technical workflow of the proposed system, illustrating interactions between the decentralized application, smart contracts, IPFS, and the TRON blockchain.

Although Figure 1 presents the technical workflow of the system, different user roles interact with the platform through the same decentralized web application. Authorized issuing entities are responsible for certificate issuance, while students and third-party users can publicly verify certificates. Access control and role differentiation are enforced at the smart contract level.

Certificate validation can be performed publicly through the developed web interface, as illustrated in Figure 2. In this interface, the user provides the cryptographic identifier (hash) of the certificate and the issuer's address, enabling direct consultation of records stored on the TRON blockchain. The system automatically returns the verification result, indicating the existence and authenticity of the certificate, without the need for intermediaries or access to centralized databases. This approach enhances the transparency and reliability of the validation process while facilitating use by different user profiles.

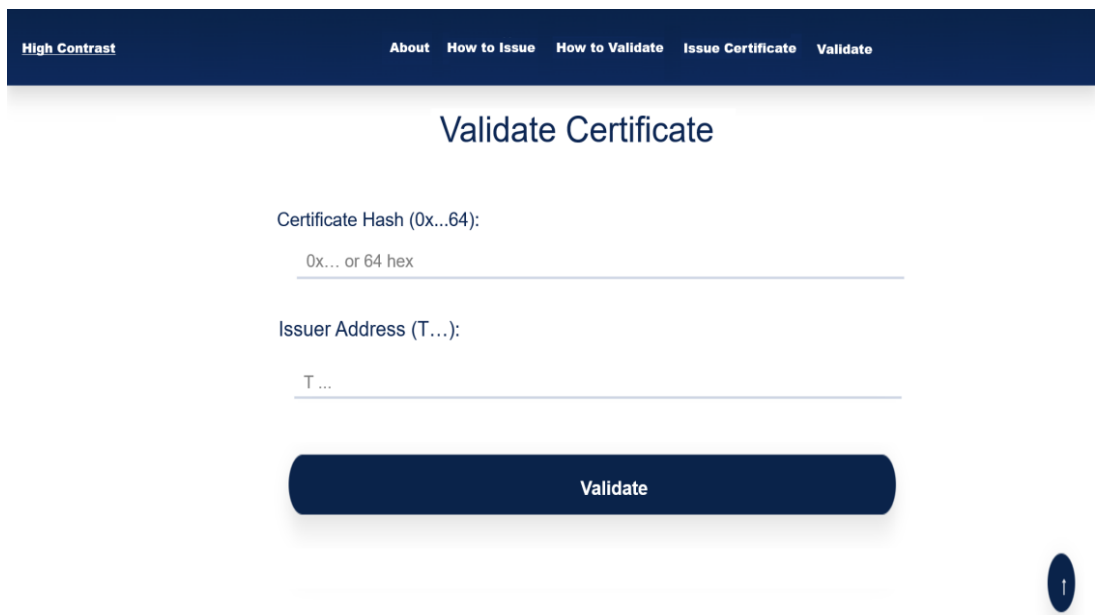

Figure 2. Web interface for academic certificate validation

## 2.5. Decentralized Storage with IPFS

IPFS was used to store certificate metadata, reducing the amount of data recorded directly on the blockchain while preserving integrity and verifiability. Each metadata file stored on IPFS is identified by a unique content-based hash, which is linked to the corresponding blockchain record.

## 2.6. Tests and Validation

System validation was carried out through different types of tests. Functional tests verified the correct operation of the certificate issuance, registration, and validation workflows. Security tests evaluated the resilience of the smart contracts against common vulnerabilities, such as fraudulent issuance, certificate duplication, and the submission of invalid parameters.

In addition, performance tests were conducted to analyse aspects such as transaction confirmation time and operational costs on the TRON network. Finally, system usability was assessed through the application of the System Usability Scale (SUS) questionnaire, completed by participants with experience in technology, allowing the measurement of user acceptance and ease of use of the developed application.

### 2.6.1. Functional Tests

The main system workflows, certificate issuance, registration, and validation, operated as expected, as summarized in Table 1.

Table 1. Smart contract functionalities and corresponding actions

| Contract Function | Available | Action Description |
| --- | --- | --- |
| Register certificate | Yes | Registration of the main certificate data (hash, identifier, dates, and metadata) on the TRON blockchain |
| Validate certificate | Yes | Verification of the existence and authenticity of the certificate using the cryptographic hash and the issuer's address |

### 2.6.2. Security Tests

Security tests were conducted to evaluate the resilience of the smart contracts against common vulnerabilities as well as potential misuse scenarios. The main tests performed and their corresponding results are summarized in Table 2.

International Journal of Computer Science & Information Technology (IJCSIT) Vol 17, No 6, December 2025

Table 2. Security tests applied to the smart contracts

| Type/Test | Item | Description | Result | Observation |
|---|---|---|---|---|
| Critical point | Replay attacks | Reuse of the same transaction with identical data | Not susceptible | It is not possible to execute multiple transactions with the same parameters |
| Critical point | Overflow/ Underflow | Variable overflow or underflow in counting operations | Not susceptible | Transactions are reverted when variable limits are exceeded |
| Simulated scenario | Fraudulent issuance | Unauthorized address attempting to issue a certificate | Transaction rejected | Certificate issuance is allowed only for previously authorized addresses |
| Simulated scenario | Certificate duplication | Attempt to issue an already registered certificate | Transaction rejected | The contract prevents the registration of duplicate certificates |
| Simulated scenario | Invalid parameters (TronWeb) | Submission of incomplete or malformed data | Transaction rejected | The transaction is not processed until valid parameters are provided |

### 2.6.3. Performance Tests

The performance evaluation aimed to assess the suitability of the TRON network for the context of the proposed application, considering aspects such as transaction confirmation latency and operational costs. To this end, public metrics of the TRON network were analyzed using the TokenView tool, which provides up-to-date information on blocks, transactions, and overall blockchain activity [16]. Although the tool does not report latency metrics in milliseconds, network performance could be inferred from the interval between recently mined blocks, which proved to be compatible with academic certificate issuance and verification applications.

### 2.6.4. Usability Tests

System usability was evaluated using the System Usability Scale (SUS) questionnaire. The instrument was applied to 12 participants with experience in technology after interacting with the certificate issuance and validation functionalities. The system achieved an average SUS score of 76.67, which indicates good usability and positive user acceptance according to the standard interpretation of the scale. These results suggest that the developed interface is suitable for the target audience and does not impose significant usability barriers.

## 3. RESULTS AND DISCUSSION

The experimental evaluation of the proposed system demonstrates that the adopted architecture is suitable for decentralized academic certificate issuance and verification. The results obtained from functional, security, performance, and usability tests provide consistent evidence of the system's technical feasibility and operational robustness.

From a functional perspective, all core workflows: certificate issuance, registration, and validation, operated as expected, as summarized in Table 1. The smart contracts correctly enforced access control policies, allowing only authorized addresses to issue certificates, while enabling public verification through cryptographic hashes. These results confirm that the proposed approach effectively supports decentralized certificate management without reliance on centralized databases.

Security tests further indicate that the implemented smart contracts are resilient to common vulnerabilities and misuse scenarios. As shown in Table 2, the system was not susceptible to replay attacks or integer overflow/underflow issues, and it successfully prevented unauthorized certificate issuance and duplication attempts. These findings are aligned with prior studies that emphasize the importance of access control and immutability in blockchain-based credential systems, reinforcing the trustworthiness of the proposed solution.

Regarding performance, the TRON network demonstrated low confirmation latency and negligible transaction costs during the execution of the tests. Although precise latency measurements in milliseconds were not available through the TokenView platform, the observed block intervals suggest effective confirmation times between approximately 3 and 6 seconds, which are adequate for academic certification use cases in educational environments. When compared conceptually to Ethereum-based solutions reported in the literature, which often face higher transaction costs and latency under increased load, the results highlight the advantages of TRON for scalable and cost-sensitive educational applications.

Usability evaluation using the System Usability Scale (SUS) resulted in an average score of 76.67, indicating good usability and positive user acceptance. This outcome suggests that the decentralized nature of the system does not impose significant usability barriers, even for users without deep technical knowledge. The combination of blockchain transparency with a user-friendly web interface represents an important factor for real-world adoption in academic environments.

Overall, the experimental results demonstrate that the proposed system achieves a balanced combination of security, efficiency, and usability. In contrast to many existing approaches that focus primarily on conceptual designs or single aspects of blockchain adoption, this work provides an integrated and experimentally validated solution, reinforcing the potential of TRON-based infrastructures for decentralized academic certification.

## 4. CONCLUSIONS

This work demonstrated the feasibility of using blockchain technology for the decentralized issuance and verification of academic certificates through the development of a functional, secure, and low-operational-cost system based on the TRON network. The proposed solution contributes to mitigating forgery risks, enhances transparency in the verification process, and reduces reliance on intermediaries, thereby strengthening institutional trust in the educational context.

The developed architecture, comprising smart contracts, a decentralized web application, and distributed storage via IPFS, proved to be well suited to the proposed scenario, as evidenced by the functional, security, performance, and usability tests conducted. The results indicate that the TRON network offers technical characteristics compatible with decentralized educational applications that require scalability, efficiency, and low transaction costs, as demonstrated by the experimental evaluations presented and discussed in this work.

Regarding limitations, the solution requires further evolution to support multiple issuing institutions, integration with legacy academic systems, and the definition of institutional and legal guidelines for the formal adoption of blockchain-based academic certificates. These aspects represent opportunities for future work, including expanding system interoperability and evaluating its adoption in real educational environments.

In this way, this study contributes to the advancement of digital transformation in education by presenting a viable decentralized solution for academic certification and by establishing relevant foundations for future research and development in the field of decentralized technologies applied to educational contexts.

## ACKNOWLEDGEMENTS

The authors would like to thank the Federal Institute of Education, Science, and Technology (IFSP) for the financial and institutional support provided through the Institutional Scientific Initiation Scholarship Program (PIBIFSP).

**AUTHORS**

**Ana Julia Evangelista Andrade** is currently an undergraduate student in Systems Analysis and Development at the Federal Institute of Education, Science, and Technology of São Paulo (IFSP), Brazil, where they are also supported by the Institutional Scientific Initiation Scholarship Program (PIBIFSP). Their research interests include blockchain technology, decentralized systems, smart contracts, and distributed applications for educational use cases. They have experience in the development and evaluation of decentralized applications (dApps) for academic certification and verification. 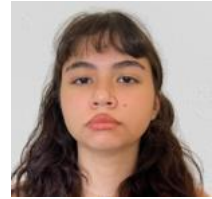

**Flavio Cezar Amate** is an Associate Professor and Researcher in the Systems Analysis and Development program at the Federal Institute of Education, Science, and Technology of São Paulo (IFSP), Brazil. Their research focuses on software engineering, distributed systems, and the application of emerging technologies such as blockchain in educational and institutional contexts. They have supervised research projects on decentralized applications, secure system architectures, and usability evaluation in educational technology. 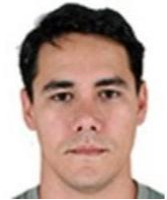